\documentclass[twocolumn]{aastex631}
\usepackage{mathrsfs}
\usepackage{amsmath}

\newcommand{\gsim}{\raisebox{-0.13cm}{~\shortstack{$>$ \\[-0.07cm]
      $\sim$}}~} 
\shorttitle{SADs in an extensive fan}
\shortauthors{Xie et al.}
\graphicspath{{./}{figures/}}
\begin{document}
\title{Supra-arcade downflows in an extensive fan associated with a giant quiescent solar filament eruption}
\author{Xiaoyan Xie}\thanks{E-mail: xiaoyan.xie@cfa.harvard.edu}
\affiliation{Harvard-Smithsonian Center for Astrophysics, 60 Garden Street, Cambridge, MA 02138, USA}

\author{Katharine K. Reeves}
\affiliation{Harvard-Smithsonian Center for Astrophysics, 60 Garden Street, Cambridge, MA 02138, USA}

\author{Chengcai Shen}
\affiliation{Harvard-Smithsonian Center for Astrophysics, 60 Garden Street, Cambridge, MA 02138, USA}

\author{Nishu Karna}
\affiliation{Harvard-Smithsonian Center for Astrophysics, 60 Garden Street, Cambridge, MA 02138, USA}

\author{Yan Xu}
\affiliation{Center for Solar Terrestrial Research, New Jersey Institute of Technology, University Heights, Newark, NJ 07102, USA}

\author{Christopher S. Moore}
\affiliation{Harvard-Smithsonian Center for Astrophysics, 60 Garden Street, Cambridge, MA 02138, USA}

\author{Crisel Suarez}
\affiliation{Harvard-Smithsonian Center for Astrophysics, 60 Garden Street, Cambridge, MA 02138, USA}

\author{Ritesh Patel}
\affiliation{Southwest Research Institute, 1301 Walnut St Suite 400, Boulder, CO 80302, USA}

\author{Daniel B. Seaton}
\affiliation{Southwest Research Institute, 1301 Walnut St Suite 400, Boulder, CO 80302, USA}



\begin{abstract}
We investigate the aftermath of a giant quiescent solar filament eruption on December 24, 2023. One feature of the eruption is an extensive fan above the filament channel that is about three times as wide as similar structures that appear above active regions (ARs) during solar flares. The fan contains numerous supra-arcade downflows (SADs), and we investigate the largest SADs with continuous Hinode X-ray Telescope (XRT) observations. The measured maximum width of the SADs in this event is at least three times the maximum width of SADs observed in AR flares, whereas the velocities of the largest SADs are similar to the typical values of AR SADs. The kinetic characteristics of the largest SADs observed in this event align with previous model predictions, where SADs originate from the non-linear development of Rayleigh–Taylor type instabilities. In this scenario, the larger system size allows the existence of larger-scale instabilities, while the development of the velocities of these instabilities is expected to be independent of the system size. Compared to AR flares, the temperature and emission measure in this event are lower, and there is less overall radiation, resulting in no evident Geostationary Operational Environmental Satellite (GOES) signature. Similar to those in AR flares, SADs show lower temperatures compared to the surrounding fan plasma. Our observations show that SADs are present in a wide variety of eruptions. The reconnection mechanisms present in quiescent filament eruptions are similar to those driving more compact eruptions originating from ARs. 
\end{abstract}

\section{Introduction} \label{sec:intro}
Supra-arcade fans provide pivotal information for understanding energy release in solar eruptions.  These structures consist of the bottom of a hot plasma sheet with temperature $\gsim$ 10 MK above flare loops \citep{2005ApJ...622.1251L,2011ApJ...727L..52R,2018ApJ...854..122W,2022MNRAS.509..406X,2024FrASS..1183746X}, which are observed from a face-on viewing perspective \citep{2017ApJ...836...55R,2019MNRAS.489.3183C,2020ApJ...898...88X,2021A&A...653A..51L}. Fans are full of dynamical structures among which the most prominent are manifested as underdense structures called supra-arcade downflows (SADs, \citealp{2003SoPh..217..247I,2012ApJ...747L..40S, 2013ApJ...766...39M}), moving sunwards with widths and velocities mainly lying in a range of 2 to 8~Mm and 20 to 400 km~s$^{-1}$, respectively \citep{2011ApJ...730...98S,2022ApJ...933...15X}. SADs have been observed in multiple-wavelengths observations, including soft X-ray \citep{1999ApJ...519L..93M,2000SoPh..195..381M,2011ApJ...730...98S}, extreme ultraviolet (EUV, \citealp{2003SoPh..217..247I,2011ApJ...742...92W,2013ApJ...767..168L,2017A&A...606A..84C,2025ApJ...986L..16F}), and white light \citep{2002ApJ...579..874S,2011ApJ...730...98S} and show correlations with nonthermal bursts of hard X-rays and microwaves \citep{2004ApJ...605L..77A,2007A&A...475..333K}. Studies have shown that SADs contribute to plasma heating in the fan by adiabatic compression \citep{2017ApJ...836...55R,2020ApJ...898...88X,2021ApJ...915..124L,2023ApJ...942...28X} and play a role in particle acceleration and transport in the regions above the flare loop-tops \citep{2025arXiv250812990L}.

Solar filaments are relatively cool ($\sim10^4$~K) and dense structures with number densities between 10$^9$ to 10$^{11}$ cm$^{-3}$ suspended in the hot corona ($\sim$1~MK) and appear darker than the surrounding plasma \citep{1985SoPh..100..415H,2005ApJ...629.1122D,2014LRSP...11....1P,2018LRSP...15....7G, 2006A&A...449L..17I,2006AdSpR..38..466Z,2019ApJ...884..157Z,2021SoPh..296..119E,2024ApJ...974..205S}. Filaments are also called prominences when protruding from the solar limb, where they appear bright in H${\alpha}$ emission against the dark background \citep{2015ApJ...807..144S}.
Large-scale filaments ($>$ 600 Mm. e.g., \citealp{2005ASPC..346..201A,2016A&A...589A..84K,2018A&A...611A..64D}) can exist in some quiet-Sun regions, and their eruptions have flare-like qualities \citep{1995JGR...100.3473H} providing conditions for investigating SADs in a larger system size.

Flux rope models representing filament eruptions have predicted large-scale reconnection current regions \citep[e.g.,][]{2004ApJ...612..519V,2017A&A...604L...7M,2018LRSP...15....7G,2019ApJ...887..103R}, where flare fan structures are expected to form. Some of these fan structures associated with filament/prominence eruptions have been observed in soft-X rays \citep{1993GeoRL..20.2785H,1998A&A...336..753K,2002ApJ...579L..45Y} and EUV \citep[e.g.,][]{2022ApJ...937L..10D}. However, there have been few reports that deliberately investigate fans/SADs associated with the eruptions of quiescent  filaments \citep{2010SSRv..151..333M,2015ApJS..221...33H,2015SoPh..290.1703M,2018ApJ...862L..23X}. 
The properties of plasma beneath erupted quiescent filaments and their dynamic behavior related to magnetic reconnection remain poorly understood.  
On December 24, 2023, the X-ray Telescope (XRT) onboard Hinode observed a fanlike structure extending across over 0.85 R$_\odot$ (solar radius) above the arcade of loops in the aftermath of a giant quiescent solar filament eruption on the northwest quadrant of the Sun (see Figures~\ref{filament} and~\ref{gen_evolution}). Inside this fan, low-intensity structures traveling sunwards were observed. No evident Geostationary Operational Environmental Satellite (GOES) X-ray flux increase was identified to be associated with this event. As this giant fanlike structure and the low-intensity structures traveling sunwards within it exhibit similar morphologies and geometry to supra-arcade fans and SADs observed during some AR flares, we refer to the fanlike structure as a supra-arcade fan in our study hereafter.

In this paper, we investigate the fan above the arcade with particular attention to the SADs in the aftermath of a giant quiescent solar filament eruption. We do not investigate details of filament eruption and only look into the filament evolution to provide general eruption context. The large width of fan above the arcade in this event ($\sim$3 times the width of AR flare fans) is rarely reported in the literature, making the investigation of the extensive fan in this event unique. We note that due to the lack of continuous XRT observations of some SADs in this event, the data is not suitable for performing statistical studies and providing the general characteristics of the SADs. In this work, we study the kinetic and thermal characteristics of the largest SADs with continuous XRT observations and compare them with the characteristics of more spatially compact flare cases. Such an investigation enables us to examine the existence of large SADs in a large filament system and explore the causes that lead to their characteristics. In the next section, we will introduce the data and methods of deriving the thermal characteristics from XRT channels in this work. We will present the results and discuss our work in Sections~\ref{sec:results} and~\ref{sec:discussion}, respectively. The supplementary diagnostics are shown in the Appendix.

\section{Data and Methods}
\label{sec:methods}
The major dataset analyzed in this paper comes from soft X-ray Telescope (XRT, \citealp{2007SoPh..243...63G}) onboard Hinode. We use the Be-thin, Al-poly, and Be-med filters. For the supplemental observations providing the general context of the eruption, we check the images observed in extreme ultraviolet (EUV) 304 \AA\ onboard STEREO-Ahead
(STEREO-A) and white light of Large
Angle and Spectrometric Coronagraph \citep[LASCO;][]{Brueckner1995SoPh} onboard Solar
and Heliospheric Observatory (SOHO). The Atmospheric Imaging Assembly (AIA, \citealp{2012SoPh..275...17L}) onboard the Solar Dynamics Observatory (SDO) is used for checking the consistency between thermal diagnostics from XRT and EUV observations.

For XRT images, we first employ the deconvolve function in XRTpy \citep{Velasquez2024}~\footnote{https://xrtpy.readthedocs.io/en/stable/} to remove the blurring caused by the telescope point spread function (PSF, details about XRT PSF can be found in \citealp{2007PASJ...59S.853W, 2016AJ....152..107A}). For each XRT image, there is a corresponding ``*.qual.fits" file providing the information of image quality. Values in ``*.qual.fits" equal to one means that the corresponding image pixel was saturated and was replaced with a constant of data number (DN) in XRT level 1 data~\footnote{https://xrt.cfa.harvard.edu/resources/documents/XAG/XAG.pdf}.  The constant DN of saturated pixels in XRT level 1 images poses an issue when performing thermal diagnostics based on the ratio of DNs in filters in this work. To solve this issue, we perform image  composition by identifying the image pixels with value 1 in the ``*.qual.fits'' file and replacing these DN/s values (DN values normalized by the exposure time) by DN/s values from the closest time with a shorter exposure time where the data was not saturated. 

For deriving the temperature and emission measure (EM) of fan from XRT observations, we use the temperature\_from\_filter\_ratio function in XRTpy. The derivation assumes an isothermal temperature distributions for the object and is based on the filter-ratio method (see details in Section~5 of \citealp{2011SoPh..269..169N}).
\section{Results} 
\label{sec:results}
In this section, we first show the general eruption context and the formation of the fan. Then we investigate the kinetic characteristics of the SADs. Finally, we present the thermal diagnostics of the fan and SADs.

\subsection{General eruption context and evolution of supra-arcade fan}
\begin{figure}
\centering
\includegraphics[scale=0.49]{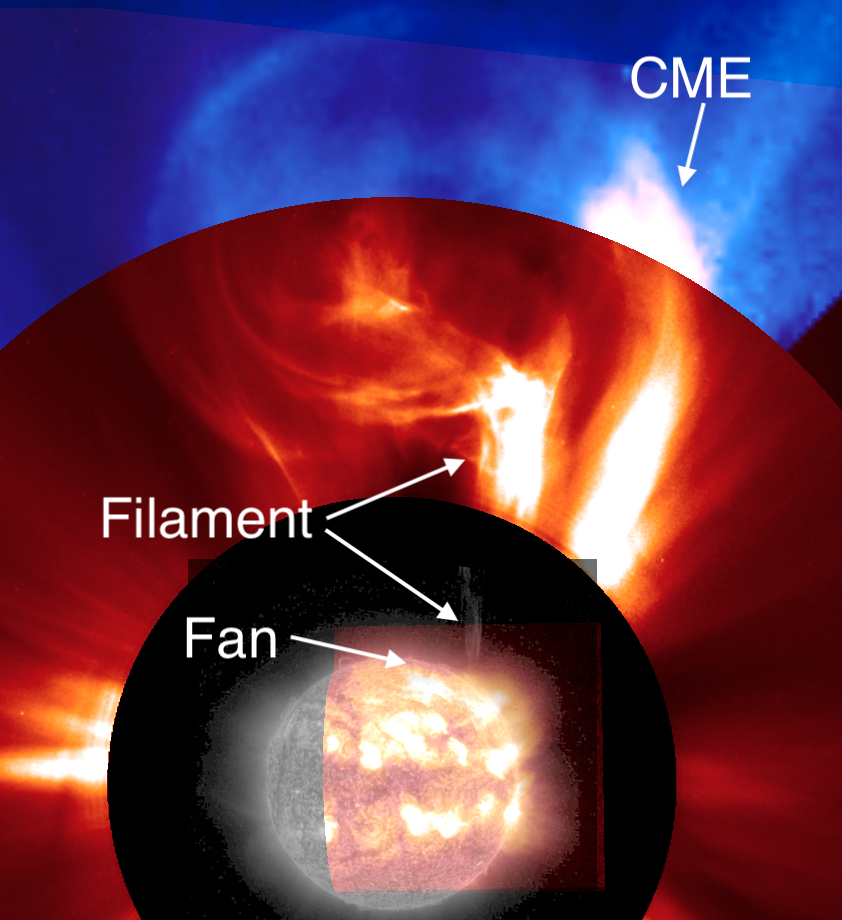}
\caption{A snapshot (made by JHelioviewer, \citealp{2017A&A...606A..10M}) displaying the general context of supra-arcade fan, quiescent filament, and the accompanied coronal mass ejection (CME) at 14:30 UT. The channels used anti-sunwards are STEREO A/304 \AA\ (gray), XRT/Be-thin (orange), LASCO/C2 (red), and LASCO/C3, respectively. The corresponding animation from 10:00 UT to 20:05 UT is available online.}
\label{filament}
\end{figure}
\begin{figure*}[ht!]
\centering
\includegraphics[scale=0.5]{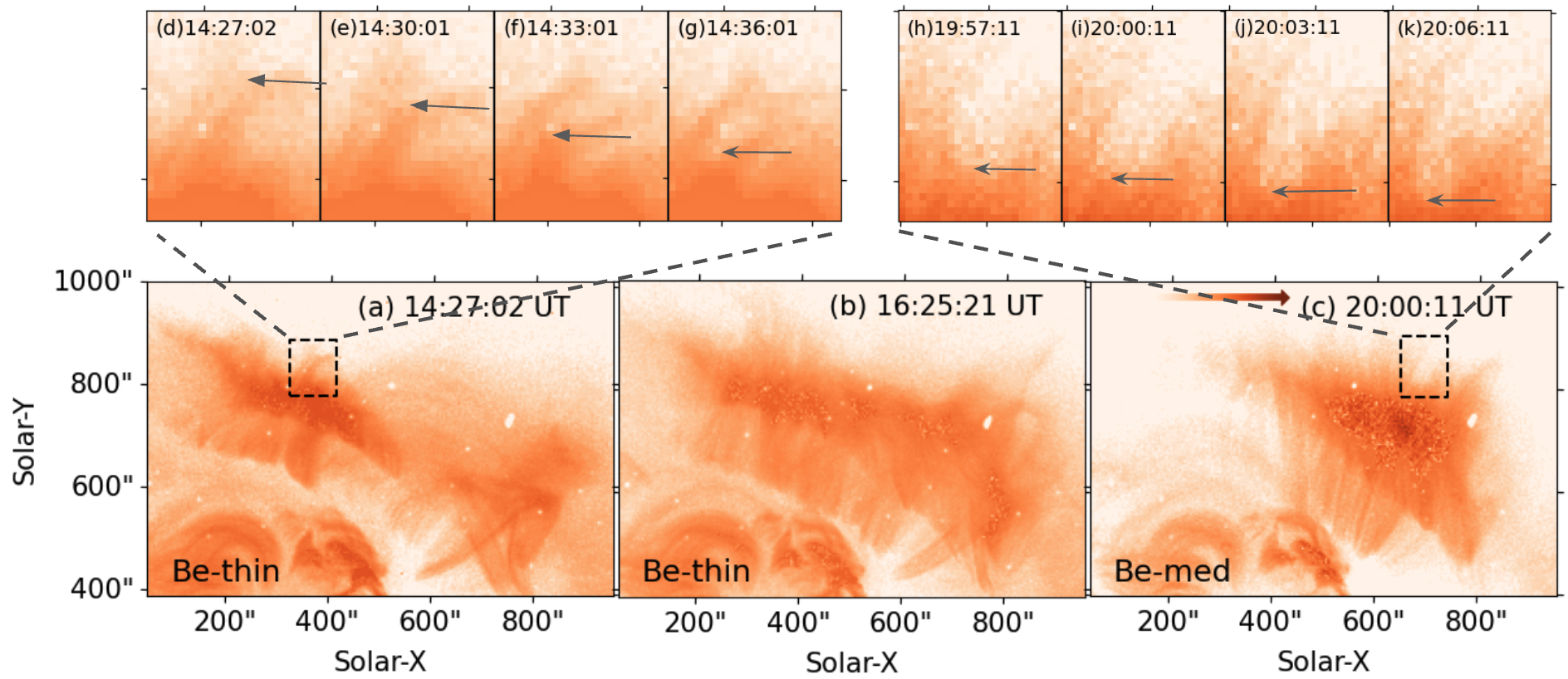}
\caption{Evolution of the supra-arcade fan and downflows observed in XRT/Be-thin and Be-med where the arrow in panel (c) indicates the color change for intensity increase. The
corresponding animation from 13:10 UT to 21:59 UT is available online.}
\label{gen_evolution}
\end{figure*}

 
On December 24, 2023, a giant quiescent solar filament erupted in the northwest quadrant of the Sun. The rise of filament material starts around 10:00 UT, and the filament fully erupts starting at around 12:10 UT (see the animation of Figure~\ref{filament}). The corresponding ribbon brightening became noticeable in STEREO A 304 \AA\ from 13:03 UT. At 13:33 UT, the accompanying coronal mass ejection (CME) is observed in LASCO/C2, and the supra-arcade fan starts to form, as shown in the observation of XRT/Be-thin (see the animation of Figure~\ref{gen_evolution}). The three components of the CME (i.e., bright core, dark dimming, and bright front), where the filament is the core, are 
discernible at 14:30 UT (Figure~\ref{filament}). The sickle-shape filament continues propagating and is  recognizable at a distance $>$ 15$R_\odot$ from the solar surface (see animation of Figure~\ref{filament}).

Figure~\ref{gen_evolution} (and its animation) shows the evolution of the turbulent supra-arcade fan area that has low-intensity supra-arcade downflows (SADs) traveling through it at multiple locations observed in XRT. The fan becomes visible and clear from 14:00 UT, with the first noticeable SAD traveling through the fan at 14:21 UT. The formation of the arcade initially starts from the side closer to the solar center and evolves towards solar western limb. At 16:25 UT, an arcade with an extensive fan above it across 0.85$R_\odot$ (solar radius) can be seen (Figure~\ref{gen_evolution}b). The arcades gradually fade and disappear starting from the left side and moving toward the right starting at 16:30 UT. The right portion of the arcade and fan is still visible at 21:00 UT with a noticeable SAD visible around 20:00 UT in the XRT observations.
\subsection{Kinetic characteristics of supra-arcade downflows} 
\begin{figure*}
\centering
\includegraphics[scale=0.68]{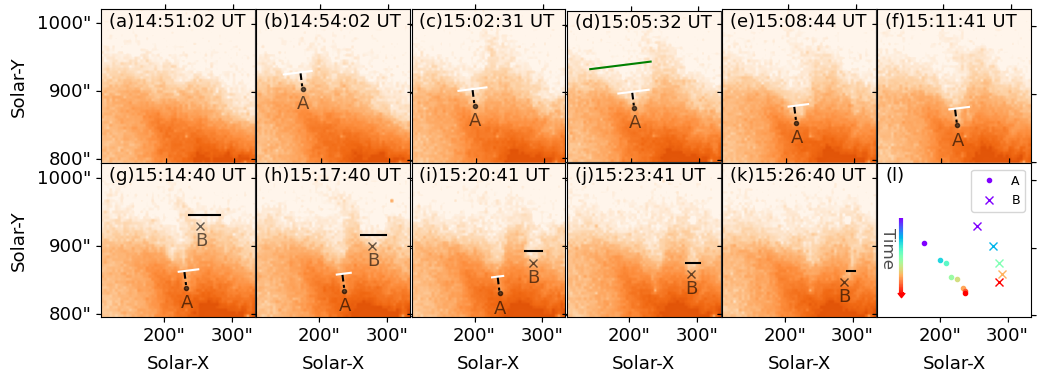}
\caption{Evolution of SADs A and B where dot and cross indicates the head of SADs observed in XRT/Be-thin (a-k). White and black solid lines indicate the corresponding SAD widths shown in Figure~\ref{wv}. Dashed line indicates the distance that is 6 pixels away from the head of SAD A. Green line in panel (d) indicates twice the width of white line at the moment. We plot positions of SAD heads at different times in panel (l). Note that time difference between panels (b) and (c) is 509~s and the rest is about 180~s.}
\label{flow_evolution}
\end{figure*}
We select two largest SADs, labeled SADs A and B, with continuous XRT observations to study the kinetic characteristics of them. Figure~\ref{flow_evolution}(a)-(k) shows the evolution of SADs A and B where the SAD heads are indicated by a dot and a cross, respectively. SAD A appears at 14:54 UT in XRT observation (Figure~\ref{flow_evolution}b) and moves sunwards with the SAD widening from 14:54 UT to 15:08 UT. Part of SAD A, particularly the portion close to the head, shrinks from 15:08 UT. On the other hand, SAD B, which appears later, shows a consistent shrinking pattern from its initial appearance at 15:14 UT (Figures~\ref{flow_evolution}g-k). We plot the positions of the SAD heads at different times in panel (l) to display the movement tracks of the SADs. SADs A and B move along arched paths which show similar features, moving westward initially and eventually moving slightly eastwards at the end.   

\begin{figure*}[ht!]
\centering
\includegraphics[scale=0.7]{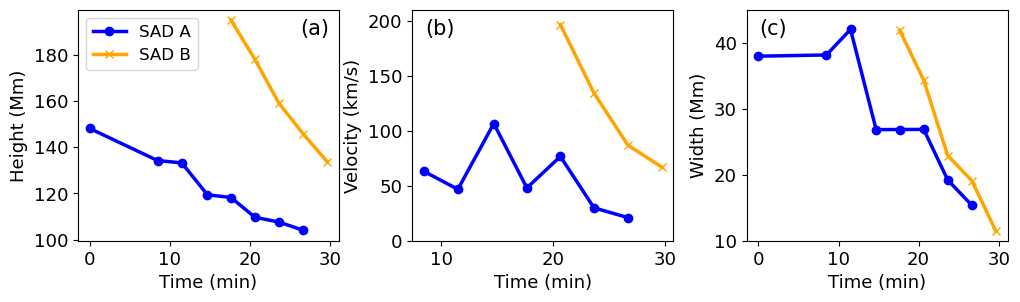}
\caption{Height, velocity, and width as a function of time for SADs A and B where we set the moment when we start tracking SAD A (i.e., 14:54:02 UT, as shown in Figure~\ref{flow_evolution}(b)) as time 0.}
\label{wv}
\end{figure*}

The kinematic characteristics of SADs A and B are shown in Figure \ref{wv} where we set the moment when we start tracking SAD A (i.e., 14:54:02 UT, as shown in Figure~\ref{flow_evolution}b) as time 0. To measure the height of the SADs from the solar surface, we first define a baseline that is parallel to the ribbons and located in the central region of the arcade loops. The black dashed line in XRT/Be-thin image in Figure~\ref{aia_xrt} marks this baseline. We then measure the distance between the heads of the SADs and the baseline, indicated by white dashed lines in Figure~\ref{aia_xrt}. In addition, as we note in Figures~\ref{gen_evolution} and~\ref{aia_xrt}, the event does not occur on the solar limb from the observation perspective of the instrument, we therefore perform de-projection calculation (see Appendix~\ref{sec:deproject}) according to the baseline position on the Sun. 

The heights of SADs A and B relative to the solar surface as a function of time are displayed in Figure~\ref{wv}(a). The initial heights of SADs A and B are 148.07 Mm and 194.80 Mm, respectively. The median and the maximum initial heights of flare SADs in the statistical results of \citet{2011ApJ...730...98S} are 82.85 Mm and and 200 Mm, respectively. The initial heights of SADs A and B are located within 10\% of the greatest initial heights in the samples in \citet{2011ApJ...730...98S}. Since Figure~8 in \citet{2011ApJ...730...98S} shows that the initial heights of flare SADs generally increase throughout the flare lifetime, we check the initial height of a SAD in the late phase, i.e., the SAD shown in Figure~\ref{gen_evolution}(h) (a different SAD from SADs A and B) to see if its height exceeds the initial heights of SADs A and B. We found that the initial height of SAD in Figure~\ref{gen_evolution}(h) is 214.66~Mm, greater than the initial heights of SADs A and B and greater than the maximum value in the statistical samples of \citet{2011ApJ...730...98S}. The change in height of SADs A and B are 43.95 Mm and 61.07 Mm, respectively, which are located within 10\% largest change in height in the samples of \citet{2011ApJ...730...98S}  (Figure~6c therein). 

Using the information of the location of the SAD heads over time, we calculate the velocities of SADs as a function of time, as shown in  Figure~\ref{wv} (b). The speeds of SADs A and B are within the typical SADs speed range of statistical samples from AR flares. SAD A has a relatively low speed, ranging from approximately 100 km/s to 25 km/s, while SAD B moves faster, with a deceleration with the speeds from about 200 km/s to 75 km/s. Both velocity evolution patterns of SADs A and B have been reported in previous publications of flare SAD samples. For example, the statistical study of \citet{2011ApJ...730...98S} shows that most SADs undergo no noticeable acceleration or a slight deceleration. On the other hand, deceleration is reported for most SADs in \citet{2021ApJ...915..124L}. 

Since SADs have irregular shapes, we need to set up a system to consistently measure the widths of SADs. We choose a line that cuts across the SAD with a distance of 6 pixels from the SAD heads and is perpendicular to the SAD travel direction to measure the SAD width. This standard for measuring SAD width is empirical, and it allows us to measure SAD width with clear intensity change of boundaries (see Appendix A). The width variation can generally represent SAD expansion and shrinkage. As shown in Figure~\ref{flow_evolution}, the dashed lines indicate the distance of 6 pixels away from the head of SAD A, and the white and black solid lines indicate the measured widths of SADs A and B along this cutting line, respectively. We then show the obtained widths of SADs A and B as a function of time in Figure~\ref{wv}(c). Consistent with the evolution displayed in Figure~\ref{flow_evolution}, we see the width of SAD A initially increases then decreases and SAD B continually decreases over time. Both width evolution patterns have been found in the AR flare SAD samples (e.g., Figure~7 in \citealp{2023MNRAS.522.4468T}).

Since for a certain SAD, the width measured in \citet{2022ApJ...933...15X} is the largest width, we also estimate the largest width for SAD A. The largest width of SAD A, as we can visually see in Figure~\ref{flow_evolution} (with the assistance of green line indicating twice the width of the white line), is usually at least twice the width we measure in Figure~\ref{flow_evolution}. The largest width for SAD A at 15:05 UT is then greater than 80~Mm, a value more than 3 times the largest widths measured in more than 600 AR flare SADs in \citet{2022ApJ...933...15X} and an order of magnitude higher than the median values in AR flare SAD samples of \citet{2022ApJ...933...15X}. 

\subsection{Thermal diagnostics for supra-arcade fan} 
\label{sec:thermal}
\begin{figure*}[ht!]
\centering
\includegraphics[scale=0.8]{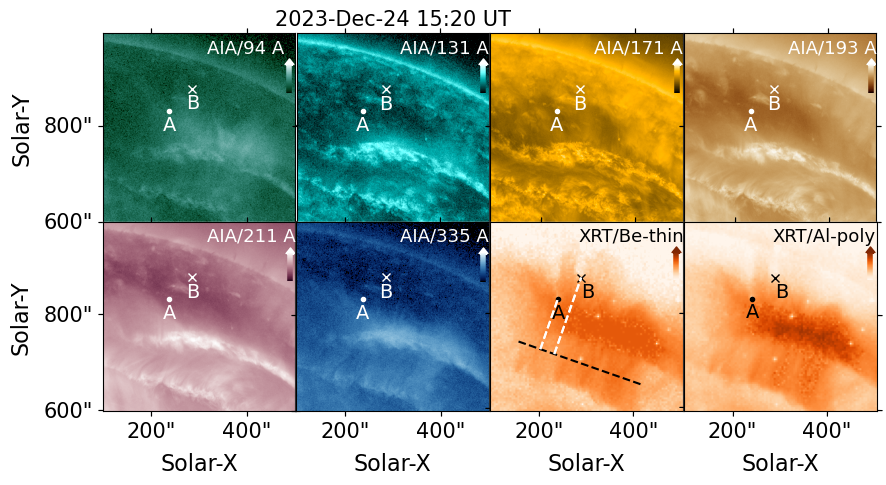}
\caption{Observations of supra-arcade fan in AIA channels and XRT filters at 2023-12-24 15:20 UT where the arrow in each panel indicates the color change for intensity increase. The locations of SADs A and B's heads identified from Figure~\ref{flow_evolution}(i) at 15:20 UT are indicated in each panel. White dashed lines indicate the height measurements for SADs before performing de-projection of plane-of-sky effect.}
\label{aia_xrt}
\end{figure*}
In this subsection, we investigate the thermal properties of supra-arcade fan and the SADs. We show the arcades and the fan at 15:20 UT in AIA and XRT/Be-thin and Al-poly observations in Figure~\ref{aia_xrt}. Putting the AIA and XRT images together illustrates that, in all iron-dominated AIA channels, the emission of the fan above the arcades shown in XRT filters can only be vaguely recognized in AIA/94 \AA. The emission of fan is mostly absent in AIA channels , making deriving fan's temperature from AIA data difficult since there is too much background (in our case, anything other than fan is the background) emission along the line of sight (LOS) contributing to the intensity in the AIA images. We therefore do not use AIA data to derive thermal properties of fan but instead use it to perform a consistency test between the XRT thermal diagnostics and the AIA observations (see Section~\ref{sec:consistency} in the Appendix)
\begin{figure*}
\centering
\includegraphics[scale=0.5]{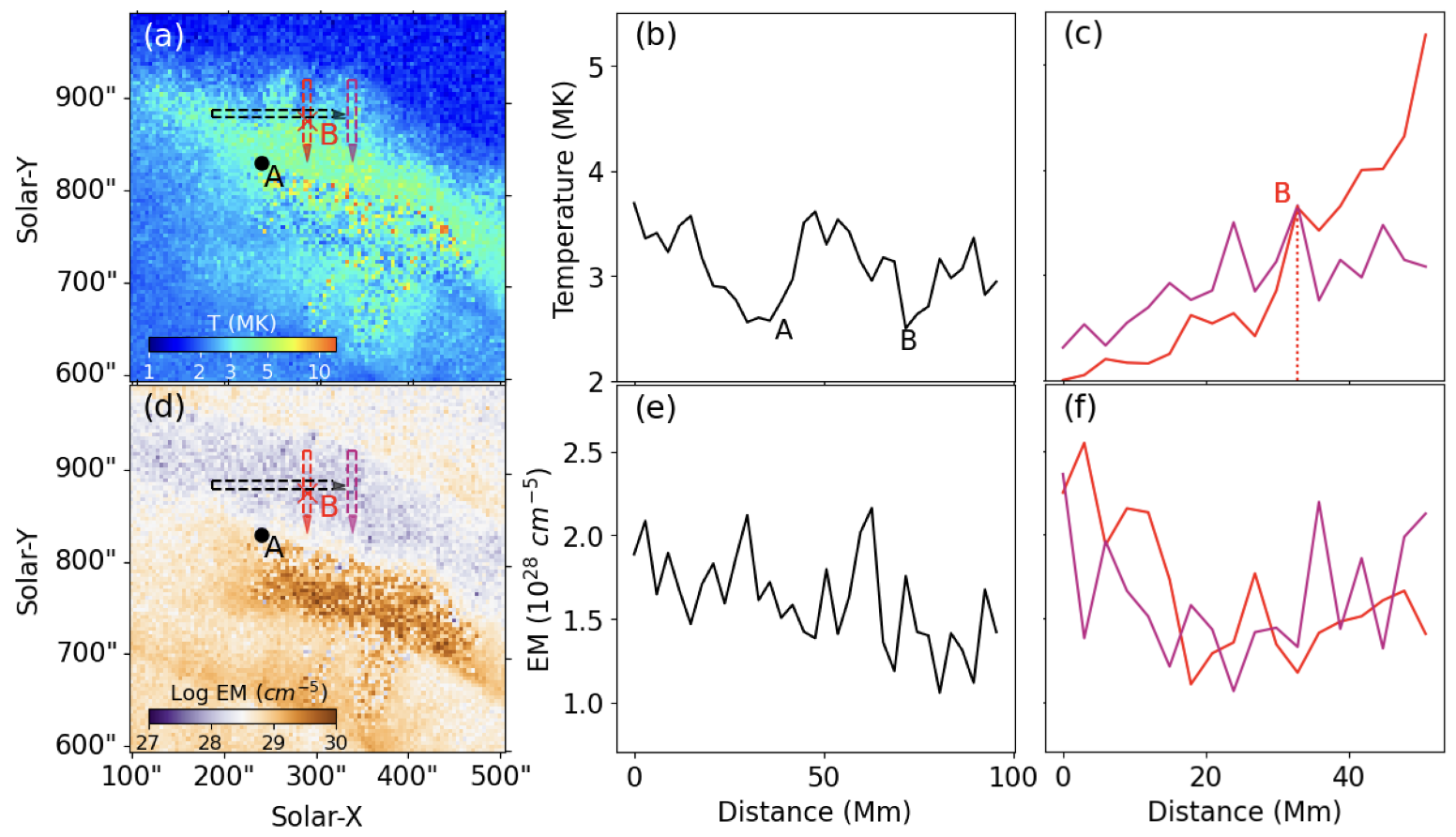}
\caption{Temperature (a) and emission measure (EM) (d) distributions at 15:20 UT derived form XRT/Be-thin and Al-poly, where dots and cross indicate the heads of SADs A and B, respectively. (b, c, e, f): Temperature and EM variations as a function of distance along the corresponding colored paths indicated in panels (a) and (d), where red dashed line indicate the locations of the head of SAD B. The temperature dips of SADs A and B are marked by ``A" and ``B" in panel (b).}
\label{tem}
\end{figure*}

We use the XRT Be-thin and Al-poly filters to derive the temperature of the fan above the arcade. The fan temperature is shown in Figure~\ref{tem}(a) where the dot and the cross mark the heads of SADs A and B identified from Figure~\ref{flow_evolution}. The temperature of the fan is about 5 MK, lower than the typical flare supra-arcade fan temperature of $\gsim$ 10 MK \citep{2014ApJ...786...95H,2017ApJ...836...55R,2023ApJ...942...28X}. Similar to AR flares \citep{2014ApJ...786...95H,2017ApJ...836...55R}, the under-temperature SADs can be noticed in Figure~\ref{tem}. 

We select three paths, as indicated in Figure~\ref{tem}(a) where the direction of the distance is indicated by the arrows, to measure the temperature variation of the fan and SADs. As shown in Figure~\ref{tem}(a), the horizontal path goes through the SADs (marked as A and B) and the vertical red path passes through the SAD B, while the dark pink path is used to monitor the average temperature outside the SADs. The temperature as a function of distance is shown in Figures~\ref{tem}(b) and (c). The black curve in Figure~\ref{tem}(b) has minima that correspond to the horizontal positions of SADs A and B, indicating that the SADs have lower temperatures than their surroundings. Both the dark pink and red curves in Figure~\ref{tem}(c), which correspond to the same colors as the paths in Figure~\ref{tem}(a), show an increase in temperature closer to the arcade top. However, the increase of temperature ahead of SAD B's head is more prominent compared to another region along the dark pink path that not pass through any visible SADs. The sharper increase of temperature ahead of SAD heads could be caused by plasma compression, same as reported in statistical samples of SADs observed during flares \citep{2017ApJ...836...55R,2023ApJ...942...28X}. 

A similar analysis but for the EM has been done and is shown in Figures~\ref{tem}(d)-(f). We note that the tenuous fan above the arcades, manifested as a low EM distribution in Figures~\ref{tem}(d), is almost an order of magnitude lower than the EM of flare fans (e.g., \citealp{2017ApJ...836...55R}). Unlike the under-dense SADs in AR flares \citep{2017ApJ...836...55R,2020ApJ...898...88X}, the curves in Figures~\ref{tem}(e) and (f) show no prominent underdense properties related to SADs, which could possibly because that the density of the whole fan itself is low and hence makes the variation of EM related to SADs hard to detect.
\section{Conclusion \& Discussion} 
\label{sec:discussion}
In this study, we investigate the kinetic and thermal characteristics of the largest SADs with continuous XRT observations in an extensive fan ($\sim$ three times width of AR flare fans) above the arcade in the aftermath of a quiescent filament eruption. We find that the maximum width of SADs is at least three times the maximum width of SADs in AR flares while the velocities are within the typical values of AR flare SADs. Such width and velocity characteristics of SADs align with the previous theory where SADs are proposed to originate from the mixture of Rayleigh–Taylor instability (RTI) and the Richtmyer–Meshkov instability (RMI)\citep{2022NatAs...6..317S}. In this theory, a larger size system (in our case, it is the width of the fan) allows for the existence of longer-wavelength instabilities, which lead to the appearance of larger SADs. The velocities of the instabilities are independent of the length scale in multiple-mode interface of a developing RTI/RMI system \citep[e.g.,][]{Alon1995PhRvL..74..534A}.  

The detected maximum initial height of SADs, which appears in the later phase than SADs A and B, in our event is 214.66 Mm, which is higher than the upper limit of SAD initial height of 200 Mm in the samples of \citet{2011ApJ...730...98S}. However, the characteristic that the SADs appear at relatively higher heights is not a unique feature compared to spatially compact AR flares. Checking the fan in an event that produces giant arches (\citealp{2015ApJ...801L...6W}, see Figure~\ref{giant_arch} in the appendix), it is plausible to reason that SADs in that event originate from higher heights that are out of SDO/AIA FOV ($\sim$208 Mm from solar surface). 
In both our event and the giant arches event, the higher initial height at which the SADs appear could be due to the higher magnetic reconnection site and instability interface. It is remarkably noteworthy that in the giant arches event, which is an AR flare, the maximum SAD width is 18.21 Mm (see Figure~\ref{giant_arch}), within the width range of the AR flare SADs sample. The comparison of maximum widths and initial heights between the giant arches event and our event further solidify our finding that maximum SAD widths are constrained by the length scale of the fan width dimension. 

The temperature of the fan in our case is lower than that in AR flares ($\sim$5 MK, compared to AR flare fan temperatures of $\gsim$10 MK), and emits low radiation that results in no recorded GOES flare \citep{2005ApJ...630.1133R}. The fan exists for more than 8 hours with SADs existing for more than 5.5 hours. The thermal characteristics of fan in our event are consistent with the conclusions in \citet{1995JGR...100.3473H}, i.e., the eruptions of quiet-Sun filaments are flare-like, but tend to be larger, slower, weaker, and cooler. Similar to AR flares, the SADs in this event show lower temperatures compared to surrounding fan plasma. However, it is hard to tell if the SADs in our case have lower EM compared to the fan because the EM of the fan itself in our case is a order of magnitude lower than that in AR flares. Similar to AR flare fans and SADs, we see a more prominent temperature increase in front of SADs than that of the counterpart fan area that do not have SADs traveling through. This temperature increase could be due to plasma compression, which is the main plasma heating mechanism of SADs identified in \citet{2017ApJ...836...55R,2020ApJ...898...88X,2023ApJ...942...28X}. The similarities of the kinetic and thermal characteristics of SADs in our event and those in AR flares indicates that there is a broader existence of SADs in various solar eruption events. In fact, recent work of \citet{2025ApJ...984L..27X} demonstrate that SADs are one aspect of turbulent flows. The complex plasma flows in the region above the flare loop-top have been reported in \citet{2013ApJ...766...39M} and \citet{2018ApJ...866...29Freed}, suggesting that turbulent flows in the region above the flare loop-top might play a critical role in energy transfer during flares. Recent high-resolution 3D MHD simulations \citep{2022NatAs...6..317S,2023ApJ...943..106S,2023ApJ...947...67R,2023ApJ...955...88Y} have shown the region above the loop-top is full of turbulent flows at different scales. \citet{2023ApJ...947...67R} illustrates that there is  more than 10\% conversion from downflow kinetic energy to turbulent energy. These findings suggest a general role that SADs could play in energy transfer in solar flares and other eruptions that do not produce flares. Also, the contribution of turbulent flows to energy transfer at different scales poses the necessity of taking the energy partition associated with turbulent flows into account when drawing the picture of flare energy transport and release in solar eruptions. It is worthwhile to have more SAD events associated with larger filament systems to extend our current work and give the general characteristics of SADs in such systems.

One limitation for current instruments with channels suitable for observing fan and SADs is usually limited by the FOV. Taking the advantage of the fact that our event  does not occur on solar limb from the viewpoint of XRT, our measurements of the highest initial appearance of SADs as 214.66 Mm confirms the existence of SADs originating from heights out of the FOV of current instruments. The inclusion of possible missing parts of SADs under current instruments could lead to the detection of SADs with longer lifetimes, wider width, and higher speeds, and provide more comprehensive understanding of SAD origin and evolution. Furthermore, as \citet{2025ApJ...984L..27X} illustrate that the anisotropic properties of turbulent flows are dependent on heights, and 3D MHD simulations of \citet{2023ApJ...943..106S} unravel the MHD instabilities in the arms of the so-called magnetic tuning fork located at the upper parts of the region above the flare loop-top. Having EUV or soft X-ray instruments with bigger FOVs than the current instruments will shine a light on understanding complex behaviors of turbulent flows and tackling the questions, such as identifying the site of magnetic reconnection and area of turbulent interface in the plasma sheet, that are vital for understanding flare energy transport and release. 
\section*{Acknowledgements}
We are grateful to Dr. Nengyi Huang for the valuable and insightful discussions. The work of X.X. and K.R. is supported by contract NNM07AB07C from NASA to SAO. X.X., K.R., and C.S. acknowledge the support from NSF grant AGS-2334929. C.S. acknowledges NASA Grant 80NSSC25K7707, 80NSSC20K131. Y.X. acknowledges NASA grant 80NSSC20K0182. The AIA data are provided courtesy of NASA/SDO and the AIA science team. Hinode is a Japanese mission developed and launched by ISAS/JAXA, with NAOJ as domestic partner and NASA and STFC (UK) as international partners. It is operated by these agencies in co-operation with ESA and NSC (Norway).  The SolarSoftWare (SSW) system is built from Yohkoh, SOHO, SDAC and Astronomy libraries and draws upon contributions from many members of those projects. We would also like to extend our thanks to the developers of XRTpy: A Python package for solar X-ray telescope data analysis (available at https://xrtpy.readthedocs.io/) for providing the tools necessary for our XRT analysis.STEREO, LASCO.

\appendix \label{sec:app}
\section{Intensity versus distance along cuts}
\label{sec:deproject}
\begin{figure}[h!]
\centering
\includegraphics[scale=0.38]{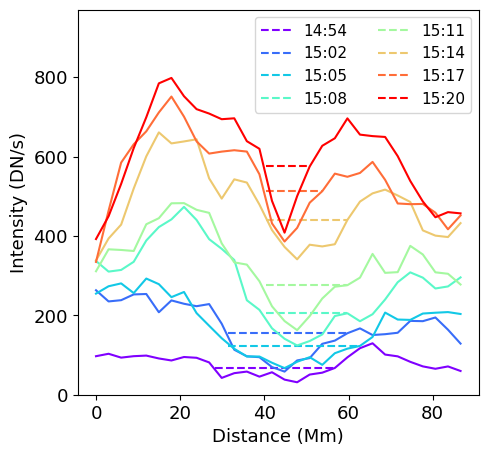}
\caption{Intensity versus distance along the extended cuts indicated in Figures~\ref{flow_evolution}(b)-(i) for SAD A. The dashed lines indicate the SAD A widths that correspond to white lines in Figures~\ref{flow_evolution}(b)-(i).}
\label{inte_cut}
\end{figure}
We choose a cut that is perpendicular to the SAD travel direction and across boundaries with a clear intensity change to allow us to measure the SAD width using the full width at half maximum (FWHM) of the intensity. After testing, we find that the cut with distance of 6 pixels away from the SAD head performs well for both SADs A and B throughout continuous evolution. Figure~\ref{inte_cut} uses SAD A as an example to show the intensity versus  distance along the extended cuts that are 6 pixels away from SAD A head at different moments. The dashed lines in Figure~\ref{inte_cut} indicate the SAD A widths that correspond to white lines in Figures~\ref{flow_evolution}(b)-(i).
\section{Illustration of De-projection}
\label{sec:deproject}
\begin{figure}[h!]
\centering
\includegraphics[scale=0.38]{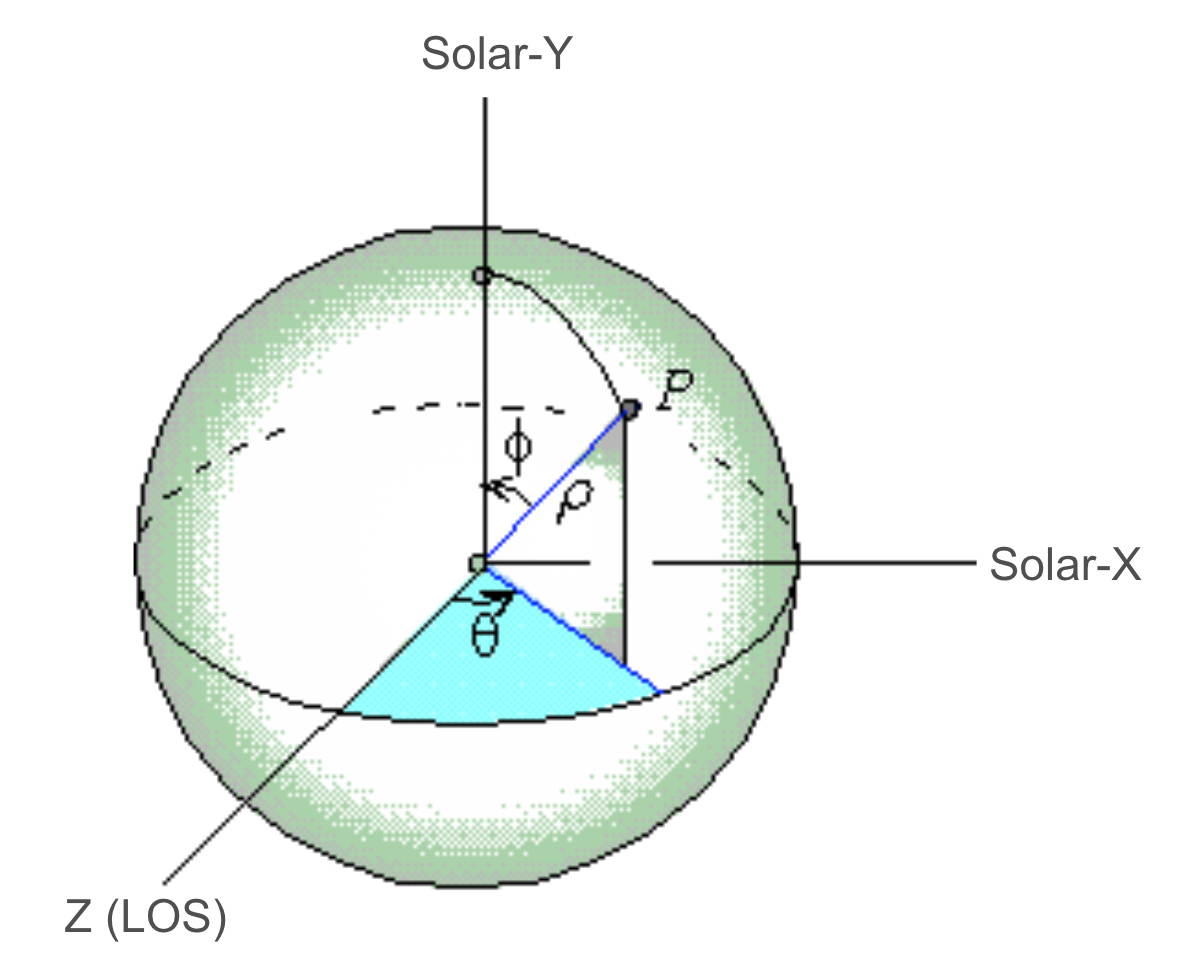}
\caption{Illustration of projection effects where solar-X and solar-Y indicate the solar image coordinate (adopted form https://math.etsu.edu/multicalc/prealpha/Chap3/Chap3-5/part2.htm).}
\label{deproject}
\end{figure}
The location of the baseline (black dashed line in Figure~\ref{aia_xrt}) on the solar image coordinate gives the information about projection effects. In Figure~\ref{deproject}, solar-X and solar-Y indicate the solar image coordinate, and P indicates the point on the baseline that is used to measure the heights on the image. The relation of the de-projected height $H$ and the measured heights $h$ from image (indicated by white dashed lines in Figure~\ref{aia_xrt}) is $H=\frac{h}{sin\phi}$. The location of P on image coordinate is 
\begin{equation}
\begin{split}
    x_{solar}&=\rho sin \phi sin \theta    \\
    y_{solar}&=\rho cos \phi
\end{split}
\end{equation}
Since P is on the solar surface, we have $\rho=R_\odot$. We therefore deduce that $H=\frac{h\sqrt{\rho^2-y_{solar}^2}}{x_{solar}}$.
\section{Consistency between XRT thermal diagnostics and AIA observations}
\label{sec:consistency}
\begin{figure*}[ht!]
\centering
\includegraphics[scale=0.46]{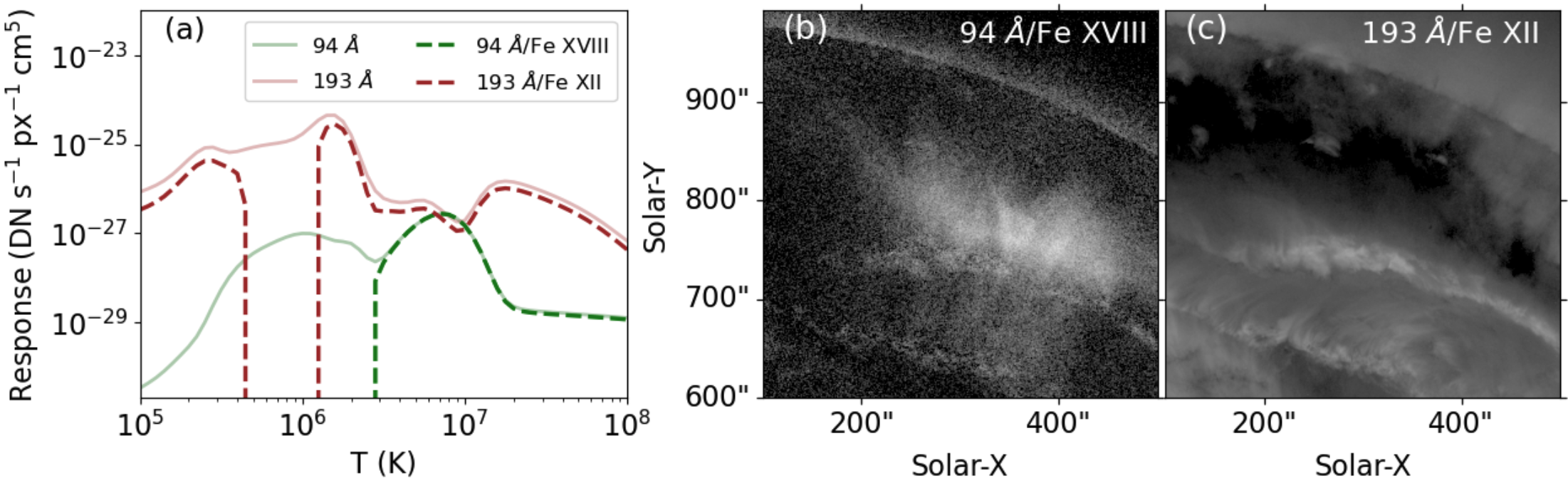}
\caption{(a): Temperature responses of AIA/94 \AA, AIA/193 \AA, AIA/94 \AA\ due to Fe XVIII line (94 \AA/Fe XVIII), and AIA/193 \AA\ due to Fe XII lines (193 \AA/Fe XII). (b): Intensity distribution of 94 \AA/Fe XVIII. (c): Intensity distribution of 193 \AA/Fe XII.}
\label{fe}
\end{figure*}
In this section, we perform a consistency test between the XRT thermal diagnostics and AIA observations. Implementing the formula provided in \citet{2013A&A...558A..73D} for estimating AIA/94 \AA\ due to Fe XVIII line (94 \AA/Fe XVIII) using AIA data, we obtain temperature response and intensity distribution for 94 \AA/Fe XVIII in Figures~\ref{fe}(a) and (b), respectively. As informed from Figures~\ref{fe}(a), compared to 94~\AA, 94 \AA/Fe XVIII removes the temperature response under $\sim$2 MK, therefore highlighting emission responding to temperature around 4 MK to 9 MK. Because the temperature derived from the XRT filters is $\sim$5 MK, we speculate that the emission of fan in 94 \AA/Fe XVIII would be more noticeable than in AIA/94 \AA. Indeed, the intensity distribution of 94 \AA/Fe XVIII (Figure~\ref{fe}b) displays more noticeable fan emission than that of 94 \AA. On the other hand, as indicated in Figure~\ref{fe}(a), compared to 193 \AA, 193 \AA/Fe XII removes the temperature response around 0.4 MK to 1 MK, plasma intensity that responds to temperature between around 4 MK to 9 MK (groove-shape in 193 \AA/Fe XII) is expected to show more noticeable noticeable dark background in 193 \AA/Fe XII than in 193 \AA. We indeed see more noticeable dark background in contrast to loops in 193 \AA/Fe XII than in 193 \AA. 

\begin{figure*}
\centering
\includegraphics[scale=0.64]{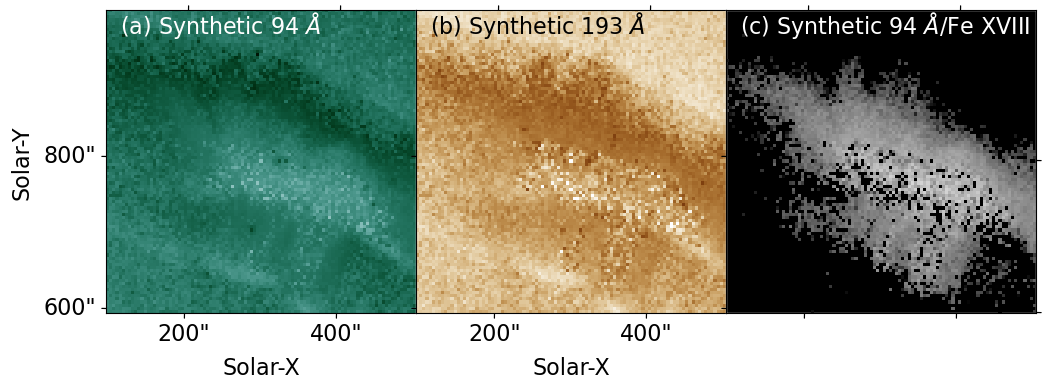}
\caption{Synthetic images from XRT-derived EMs.}
\label{syn}
\end{figure*}

In addition, we use XRT-derived temperature and EM to perform synthetic AIA intensity by employing the formula $ I = \int R(T) \times {\rm DEM}(T) dT$ \citep{2012SoPh..275...41B} where $R(T)$ is the temperature response for AIA channels. In our case with isothermal temperature, the formula can be further simplified to $I = R(T_{iso}) EM(T_{iso})$. Since our intention is to check if using XRT-derived temperature and EM could generate synthetic AIA images showing the relative emission patterns between the arcades and the fan as in AIA observations, we do not take care of the difference of pixel size and cross-calibration factors into account in this study. For the same reason and keeping in mind that the temperature for synthesis is isothermal, we do not expect the details of the fan in synthetic images to be the same as in observations. Similar to Figure~\ref{aia_xrt}), Figure~\ref{syn} display the emission of fan in synthetic 94 \AA\ and absence of fan emission compared to the arcades in synthetic 193 \AA. Because the temperature of fan is 3-5 MK, synthetic 94 \AA/Fe XVIII (Figure~\ref{syn}(c)) has more prominent fan emission by contrast to the arcades than in synthetic 94 \AA. 

The combination of low EM and temperature of the fan (compared to flare fan) in our event results in faint emission of fan in AIA observations.  The emission enhancement in 94 \AA/Fe XVIII than in 94 \AA\ and more noticeable dark background in 193 \AA/Fe XII than in 193 \AA, and the same patterns of relative emissions between the fan and loops in synthetic AIA images from XRT-derived temperature and EM confirm the consistency between our thermal diagnostics and AIA observations. This consistency justifies the feasibility of assuming isothermal temperature when calculating temperature from XRT filters in our case, i.e., the isothermal temperature from XRT filters can represent the dominant temperature of the fan.

\section{Measurement of SAD width in giant arches event}
\begin{figure}[h!]
\centering
\includegraphics[scale=0.58]{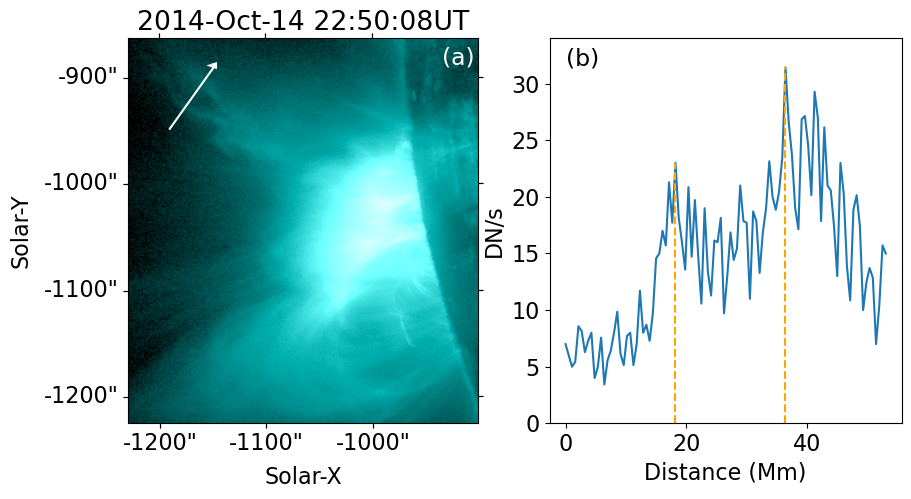}
\caption{(a) Snapshot of the giant arches event showing the SAD that has the largest width therein. Arrow indicates the direction of distance increment. (b): intensity as a function of distance along the path (white line) marked in panel (a). Orange lines indicate the range of SAD width.}
\label{giant_arch}
\end{figure}
We measure the width of the largest SAD (Figure~\ref{giant_arch}) in giant arches event (i.e., \citealp{2015ApJ...801L...6W}) of a AR flare. It shows that the maximum SAD width in giant arches event is 18.21 Mm.
\bibliography{sample631}{}
\bibliographystyle{aasjournal}



\end{document}